\documentclass{crc-book}
\usepackage{CJK} 

\usepackage{float}
\usepackage{amsthm,,amssymb,color,epsfig,graphics}
\usepackage{xcolor}
\usepackage{graphicx}
\usepackage[intlimits]{amsmath}
\usepackage{latexsym}

\newtheorem{theorem}{Theorem}
\newtheorem{lemma}{Lemma}

\newtheorem*{acknowledgments}{Acknowledgments}

\theoremstyle{definition}
\newtheorem{definition}{Definition}

\newtheorem{remark}{Remark}

\def \today{June~8, 2018}

\def \ccomma{\raise 2pt\hbox{,}} 

\def \D {\hbox{d}}
\def \sech{\mathop{\rm sech}\nolimits}
\def \coth{\mathop{\rm coth}\nolimits}
\def \tanh{\mathop{\rm tanh}\nolimits}
\def \Re  {\mathop{\rm Re}\nolimits}
\def \Im  {\mathop{\rm Im}\nolimits}
\def \arg {\mathop{\rm arg}\nolimits}
\def \mod#1{\vert #1 \vert}

\def \Degree{\mathop{\rm deg}\nolimits}
\def \Order{\mathop{\rm order}\nolimits}
\def \jmax{J}

\def\barp{\bar p}
\def\barq{\bar q}
\def\barr{\bar r}

\def\GLA{A}
\def\GLa{a}
\def\GLac{\bar a}
\def\barC{\overline{\mathbb{C}}} 

\def\dloga{U} 

\def\CA{c_a} 

\def\Pn     {{\rm Pn}}

\usepackage{geometry,multicol}
\newdimen\stockheight
\newdimen\stockwidth
\stockwidth\paperwidth
\usepackage[center,cam]{crop}
\begin{document}

%
\renewcommand{\evenhead}{Robert Conte, Tuen Wai Ng, Chengfa Wu}
\renewcommand{\oddhead}{Meromorphic solutions of differential equations}

%
\thispagestyle{empty}

\Name{Singularity methods for meromorphic solutions
of differential equations\footnote{\textit{Nonlinear systems and their remarkable mathematical structures},
ed.~N.~Euler (CRC Press, Boca Raton, 2018).}}

\Author{Robert Conte~$^{ab}$, Tuen Wai Ng~$^{b}$ and Chengfa Wu~$^c$}
\Address{
$^a$ 
    Centre de math\'ematiques et de leurs applications  
\\ \'Ecole normale sup\'erieure de Cachan, CNRS, Universit\'e Paris-Saclay,
\\ 61, avenue du pr\'esident Wilson, F--94235 Cachan Cedex, France.
\\ Robert.Conte@cea.fr
\\[10pt]
$^b$ 
Department of Mathematics,
The University of Hong Kong,
Pokfulam Road, Hong Kong.
\\ ntw@maths.hku.hk
\\[10pt]
$^c$ 
Institute for Advanced Study,
Shenzhen University,
Nanshan district,
Shenzhen, Guangdong, China 518060.
\\ CfWu@szu.edu.cn, WuChengfa9@126.com
}

\centerline{\today}

\begin{abstract}
This chapter is mainly a tutorial introduction to methods recently developed 
in order to find all (as opposed to some) meromorphic particular solutions
of given nonintegrable, autonomous, algebraic ordinary differential equations
of any order. The examples are taken from physics and include
Kuramoto-Sivashinsky and complex Ginzburg-Landau equations.
\end{abstract}

\tableofcontents
 \noindent

\section{Introduction}

Consider an $N$-th order algebraic ordinary differential equation (ODE)
 for $u(x)$
which may or may not possess the Painlev\'e property (PP),
defined as the absence of movable critical singularities in the general solution,
a singularity being called ``critical'' if multivaluedness takes place around it.
This does not prevent the existence of particular solutions
obeying a lower order ODE with the PP.
Let us therefore address the problem to find \textit{all} 
such solutions (i.e.~without movable critical singularities)
in closed form.
We restrict here to autonomous ODEs,
i.e.~which do not depend explicitly on the independent variable $x \in \mathbb{C}$.

Moreover, 
in order to shorten the writing,
instead of the mathematically correct adjectives ``uniformizable''
(synonym of ``without movable critical singularities'')
and ``nonuniformizable'',
we will use ``singlevalued'' and ``multivalued''.
For instance, if the singularity $x=0$ is fixed
and the singularity $x=x_0$ is movable,
the expression $u(x)=\sqrt{x} (x-x_0)^{-3}$
will be called here ``singlevalued''.

Let us denote $\barC$ the analytic plane,
i.e.~the complex plane $\mathbb{C}$ compactified by addition of 
the unique point at infinity.

\begin{definition} 
A function $x \to u(x)$ from $\mathbb{C}$ to $\barC$ is called 
\textit{meromorphic} on $\mathbb{C}$ 
if its only singularities in $\mathbb{C}$ 
are poles.
Example: $1/x+\coth(x-\pi)$.
\end{definition}

Two kinds of methods will be used:
(i) those based on the structure of singularities of the solutions
(Painlev\'e analysis \cite{CMBook}),
(ii) those based on the growth of solutions near infinity 
(Nevanlinna theory \cite{LaineBook}).
We will not use here methods based on Lie symmetries.

These two methods present an important difference:
Painlev\'e analysis always considers the general solution (or particular solutions,
i.e.~obtained from the general solution by assigning numerical values
to the integration constants)
and only assumes singlevaluedness,
while Nevanlinna theory considers particular solutions (which may be the general solution)
and always assumes meromorphy on $\mathbb{C}$. 
However, the synergy of these two methods is considerable,
the main achievement being that,
under mild assumptions,
any solution meromorphic on $\mathbb{C}$ is necessarily elliptic or degenerate elliptic and hence it allows one to find {\it all} meromorphic solutions.
 
As compared with previous tutorial presentations of ours
\cite{CM2005}
\cite{CMBook} 
\cite{CM2009} 
\cite{ConteNgODE3} 
\cite{WCFThesis} 
this short paper insists on the following features.

\begin{enumerate}

\item
The criterium allowing one to conclude that any solution which is meromorphic on $\mathbb{C}$ 
is necessarily elliptic or degenerate elliptic.

\item
The advantages,
when dealing with amplitude equations 
(like complex Ginzburg-Landau or even nonlinear Schr\"odinger),
to consider the logarithmic derivative of the complex amplitude $A$
instead of the natural physical variables 
$\mod{A}^2$ (square modulus) and $\arg A$ (phase).

\item
The Hermite decomposition of elliptic or degenerate elliptic functions
as a finite sum of poles.

\end{enumerate}

First, it is important to understand the consequence of the term ``closed form''.
By the construction of the Painlev\'e school \cite{CMBook},
the closed form of
any singlevalued solution (whether particular or general) of
an $N$-th order ODE
is a finite expression built from 
\begin{enumerate}
\item
the general solution of linear ODEs of any order;

\item
the general solution of irreducible nonlinear ODEs of order one
(i.e.~elliptic functions),

\item
the general solution of irreducible nonlinear ODEs of order two
(i.e.~the six functions $\Pn$ of Painlev\'e),
three (no such function), \dots, up to $N$ included.

\end{enumerate}

As a consequence, since H\"older proved the nonexistence of an ODE
obeyed by the $\Gamma$ function,
this function can never contribute to the solution
of an autonomous ODE.

Let us now explain why there exist two privileged subsets of
singlevalued functions,
the elliptic functions and the meromorphic functions.

\subsection{First order equations and elliptic functions}

Irreducible first order ODEs are privileged in this search for 
closed form singlevalued solutions,
because order one is the smallest nonlinear order.

As proven by Lazarus Fuchs, Poincar\'e and Painlev\'e
\cite[vol II \S 141]{Valiron},
irreducible first order ODEs with the PP (autonomous and nonautonomous) 
have a general solution which is a rational function of the Weierstrass function $\wp(x,g_2,g_3)$
and its first derivative.

\begin{definition} 
A function $x \to u(x)$ from $\mathbb{C}$ to $\barC$ is called 
\textit{elliptic}
if it is doubly periodic and meromorphic on $\mathbb{C}$.
Example: $\wp'(x)+\wp(2 x)$.
\end{definition}

All the elliptic functions $f(x)$
(among them the twelve Jacobi functions)
are rational functions of $\wp(x)$ and $\wp'(x)$,
\begin{eqnarray}
& &
f(x)=R(\wp(x),\wp'(x)).
\label{eqRationalwpwpprime}
\end{eqnarray}

This function $\wp$ of Weierstrass is defined as the general solution of
the first order ODE
\begin{eqnarray}
& &
{u'}^2= 4 u^3 - g_2 u - g_3
        =4(u-e_1)(u-e_2)(u-e_3),\
\end{eqnarray}
it is doubly periodic
and its only singularities are a lattice of double poles.
Elliptic functions $f(x)$ have two successive degeneracies,
\begin{itemize}
\item
when one root $e_j$ is double ($g_2^3-27 g_3^2=0$),
degeneracy to simply periodic functions
(i.e.~rational functions of one exponential $e^{k x}$)
according to
\begin{eqnarray}
& &
\forall x,d:\
\wp(x,3 d^2,-d^3)
 = 2 d         - \frac{3 d}{2} \coth^  2  \sqrt{\frac{3 d}{2}} x,
\label{eqwpcoth}
\end{eqnarray}

\item
when the root $e_j$ is triple ($g_2=g_3=0$),
degeneracy to rational functions of $x$.

\end{itemize}

The reason why elliptic functions 
(one should say \textit{the} elliptic function because
they are all equivalent under birational transformations)
are privileged is that they are the only functions defined by
irreducible (i.e.~not linearizable) first order algebraic ODEs.
They are therefore the natural building blocks
for representing a wide class of particular solutions
of ODEs of higher order.

\subsection{Second order equations and meromorphic functions}

Irreducible second order first degree ODEs in the class
\begin{eqnarray}
& &
u''=F(u',u,x),\
\label{eqClassOrder2Degree1}
\end{eqnarray}
with $F$ rational in $u'$ and $u$, analytic in $x$,
have the property that all their movable singularities are poles
\cite{PaiActa}.
Note that, if the dependence on $u$ is algebraic,
there exist equations with movable 
isolated essential singularities in their general solution.

Therefore, elliptic functions naturally occur at first order,
and meromorphic functions at second order.

\subsection{Insufficiency of the truncation methods}

Let us now explain why the so-called ``truncation methods'' cannot achieve
our goal (to find all singlevalued solutions).

The four types of expressions considered up to now
obey the sequence of inclusions 
\begin{eqnarray}
& &  
\hbox{ multivalued }  \supset 
\hbox{ singlevalued } \supset
\hbox{ meromorphic }  \supset
\hbox{ elliptic },
\label{eqsubsets}
\end{eqnarray}
and this sequence admits an important group of invariance.

\begin{definition} 
One calls \textit{homographic (or M\"obius) transformation}
any transformation
 $u \to U$, $u=(a U+b)/(c U+d)$,
$a d - b c=1$, with $a,b,c,d$ constant.
\end{definition}

Homographic transformations are the unique bijections 
of the analytic plane $\barC$.

Each of the four above mentioned subsets
is invariant under homographies
(on the condition to consider holomorphic functions as particular
meromorphic functions,
so that for instance the transform $U$ of $u=1/x$ by $u=1/U$ is meromorphic).

Therefore, any method claiming to possibly find all singlevalued solutions
\textit{must} be invariant under homographies.

\textit{Truncation methods}\footnote{
``New expansion methods'' which are not new at all are regularly published,
see a critical account in Ref.~\cite{Kud-be-careful}.}
are the main class of methods able
to find singlevalued particular solutions.
Their common feature 
(see the summer school lecture notes \cite{CetraroConte})
is to assume the unknown solution
to belong to a given class of expressions,
for instance polynomials in $\tanh (k x)$ and $\sech (k x)$ \cite{CM1992}.
They essentially rely on 
the pioneering work of Weiss, Tabor and Carnevale \cite{WTC}.
Although very successful to find some solutions,
they miss by construction any solution outside the given class.
For instance,
if one considers the rational trigonometric function
\begin{eqnarray}
& &
u=\frac{\tanh(\xi-\xi_0)}{2+\tanh^2(\xi-\xi_0)},
\end{eqnarray}
and builds the first order ODE which it obeys
(this is a common way to construct examples),
\begin{eqnarray}
& &
{u'}^2+\left(12 u^2 - \frac{3}{2}\right) u'
 + 36 u^4 - \frac{17}{2} u^2 + \frac{1}{2}=0,
\label{eq_u_rational}
\end{eqnarray}
then no assumption $ u$ polynomial in $\tanh (k x)$ can succeed to find its solution.

One immediately notices the reason for this failure:
this is the noninvariance of these truncation methods under homographies.

It is therefore necessary to build a method invariant under homographies.

Up to now,
we do not know of a method able to find all singlevalued solutions,
but there do exist methods to possibly find all meromorphic solutions,
this is the subject of the present chapter.

\section{A simple pedagogical example}

Let us first outline the method on a very simple example.

The Kuramoto and Sivashinsky (KS) equation
\begin{eqnarray}
& &
v_t + \nu v_{xxxx} + b v_{xxx} + \mu v_{xx} + v v_x =0,\ \nu \not=0,
\label{eqKS}
\end{eqnarray}
with  $(\nu,b,\mu)$ constants in $\mathbb{R}$,\
admits a travelling wave reduction defined as
\begin{eqnarray}
& &
v(x,t)=c+u(\xi),\ \xi=x-ct,\
 \nu u''' + b u'' + \mu u' + \frac{u^2}{2} + A = 0,
\label{eqKSODE}
\end{eqnarray}
in which $A$ is an integration constant.
It has a chaotic behavior \cite{MannevilleBook},
and it depends on two dimensionless parameters,
$b^2/(\mu \nu)$ and $\nu A / \mu^3$.

It admits only eight meromorphic solutions,
which are all elliptic or degenerate of elliptic,
\begin{enumerate}

\item
one nondegenerate elliptic solution 
\cite{FournierSpiegelThual,KudryashovElliptic}
with codimension one,
\begin{eqnarray}
& &
\left\lbrace
\begin{array}{ll}
u=\displaystyle{- 60 \nu \wp' - 15 b \wp - \frac{b \mu}{4 \nu}},\
\\
b^2=16 \mu \nu,\
g_2=\displaystyle{\frac{\mu^2}{12 \nu^2}},\
g_3=\displaystyle{\frac{13 \mu^3 + \nu A}{1080 \nu^3}};
\end{array}
\right.
\label{eqKSElliptic}
\end{eqnarray}

\item
six degenerate elliptic solutions rational in one exponential 
\cite{KudryashovKSFourb} with codimension two,
\begin{eqnarray}
u & = &
 120 \nu \tau^3
 - 15 b  \tau^2
 +\left(- 30 \nu k^2 - \frac{15 (b^2-16 \mu \nu)}{4 \times 19 \nu}\right)
 \tau
\nonumber
\\
& &
 + \frac{5}{2} b k^2 - \frac{13 b^3}{32 \times 19 \nu^2}
  + \frac{7 \mu b}{4 \times 19 \nu},\
\tau  =  \frac{k}{2} \coth \frac{k}{2} (x-x_0),
\label{eqKSTrigo}
\end{eqnarray}
detailed in Table \ref{TableKS};

\begin{table}[h] 
\caption[Kuramoto-Sivashinsky equation. The six trigonometric solutions.]{
The six trigonometric solutions of KS, Eq.~(\ref{eqKSODE}).
They all have the form (\ref{eqKSTrigo}).
The last line is a degeneracy of the elliptic solution
(\ref{eqKSElliptic}).
}
\vspace{0.2truecm}
\begin{center}
\begin{tabular}{| c | c | c |}
\hline 
$b^2/(\mu\nu)$ & $\nu A/\mu^3$ & $\nu k^2/\mu$
\\ \hline \hline 
$0$ & $-4950/19^3,\ 450/19^3$ & $11/19,\ -1/19$
\\ \hline 
$144/47$ & $-1800/47^3$ & $1/47$
\\ \hline 
$256/73$ & $-4050/73^3$ & $1/73$
\\ \hline 
$16$ & $-18,\ -8$ & $1,\ -1$
\\ \hline 
\end{tabular}
\end{center}
\label{TableKS}
\end{table}

\item
one rational solution with codimension three,
\begin{eqnarray}
u & = &
 120 \nu (x-x_0)^{-3},\ b=\mu=A=0.
\label{eqKSRational}
\end{eqnarray}
\end{enumerate}

Let us prove that no more meromorphic solutions exist.

\subsection{Outline of the method}
      \label{sectionOutline_method}

The successive steps of the method, to be detailed soon, are
\begin{enumerate}

\item
Count the number of different Laurent series of $u$.

\item
Apply a useful result of Nevanlinna theory,
known as Clunie's lemma to show that $u$ has infinitely many poles if $u$ is nonrational.

If in addition the number of Laurent series is finite (which follows from the non-existence of positive Fuchs indices), one can prove that any meromorphic solution is elliptic or degenerate elliptic.

\item
If the solution is nondegenerate elliptic, 
compute its closed form, for instance by its Hermite decomposition.

\item
If the solution is rational in one exponential $e^{k x}$,
with $k$ unknown,
then its closed form is again provided by its Hermite decomposition.

\item
If the solution is rational in $x$, 
then its closed form is provided by the classical partial fraction decomposition.

\end{enumerate}

Before proceeding,
let us mention the main contributions to the method,
in chronological order:
\cite{MC2003},
\cite{Hone2005},
\cite{EremenkoKS}, 
\cite{CM2009},
\cite{ConteNgODE3}, 
\cite{DK2011method}. We would also like to reiterate that this method (originated from Eremenko's work \cite{EremenkoKS}) allows one to find {\it all} meromorphic solutions if they exist.

Let us now detail the successive steps on our pedagogical example.

\subsection{Step 1. Counting the Laurent series}
      \label{sectionCounting}

Taylor series are here excluded.

One applies the method of Kowalevski and Gambier \cite{Cargese1996Conte} 
to look for all algebraic behaviours
\begin{eqnarray}
& & u = u_0 \chi^{p},\ u_0 \not=0,\ \chi=x-x_0,
\end{eqnarray}
in which $p$ is not a positive integer.
This defines the unique dominant behaviour $p=-3, u_0=120 \nu$,
obtained by requiring the terms $\nu u'''$ and $u^2/2$ to both behave 
like the same power $\chi^q$, here $q=-6$.

Next, one must compute the Fuchs indices \cite{CMBook} of the linearized equation
near this triple pole.
This is a classical computation, which can be detailed as
\begin{eqnarray}
& &
\lim_{\chi \to 0} \chi^{-j-q} (\nu \partial_x^3 + u_0 \chi^p) \chi^{j+p}
\\
& &
= \nu (j-3)(j-4)(j-5) + 120 \nu = \nu (j+1) (j^2 -13 j + 60)
\\
& &
=\nu (j+1) \left(j-\frac{13 + i \sqrt{71}}{2}\right)
           \left(j-\frac{13 - i \sqrt{71}}{2}\right)=0.
\label{eqKSIndicial}
\end{eqnarray}
Apart from $-1$, no Fuchs index is a positive integer,
therefore the Laurent series exists,
\begin{eqnarray}
{\hskip -12.0 truemm}
& &
u = 120 \nu \chi^{-3} - 15b \chi^{-2}
        - \frac{15 (b^2-16 \mu \nu)}{4 \times 19 \nu} \chi^{-1}
        + \frac{13 (4  \mu \nu - b^2) b}{32 \times 19 \nu^2}
        + O(\chi),\
\label{eqKSODELaurent}
\end{eqnarray}
where $\chi=x-x_0$ and it is unique (no arbitrary coefficient enters the series);
consequently, for each set of arbitrary values of $\nu,b,\mu,A$, 
there exists at most one one-parameter singlevalued 
function solution of
(\ref{eqKSODE}).

\subsection{Step 2. Clunie's lemma}
\label{sectionClunie}

This is a very useful lemma in Nevanlinna theory 
to show that certain transcendental (i.e.~nonrational)
 meromorphic function has infinitely many poles.
This theory uses a specific vocabulary which we will not detail here,
referring to the book \cite{LaineBook}.
Combined with the finiteness of the number of Laurent series,
its main conclusion is that ``meromorphic implies elliptic
or degenerate elliptic''.

\begin{lemma}[Clunie's lemma] 
Let $f(z)$ be a nonrational meromorphic solution of 
\begin{equation} f^n P(z,f,f',...)=Q(z,f,f',...),
\label{eqClunie}
\end{equation} 
where 
$n$ is a nonzero positive integer, 
$P$ and $Q$ are polynomials in $f$ 
and its derivatives with meromorphic coefficients 
$\{a_\lambda|\lambda \in I\}$, such that 
for all $\lambda \in I$, $m(r,a_{\lambda})=S(r,f)$,
where $I$ is some known index set. 
If the total degree\footnote{
Defined as the global degree in all derivatives $f^{(j)}, j \ge 0$.}  
of $Q$ as a polynomial in $f$ 
and its derivatives is less than or equal to $n$, then 
\begin{equation}m(r,P(z,f,f',...))=S(r,f).\end{equation}
\label{lem2}
\end{lemma}

Equation (\ref{eqKSODE}) does fit (\ref{eqClunie}),
with $n=1$, $P=u$, $Q=-2(\nu u''' + b u'' + \mu u' + A)$,
and all its coefficients obey the smallness assumption of the lemma
because they are constant. Hence,  it follows from Clunie's lemma that $m(r,P)=m(r,u)=S(r,u)$. By the First Main Theorem of Nevanlinna theory (\cite{LaineBook}), one can deduce that $u$ must have infinitely many poles $\{z_n\}$ if $u$ is a transcendental meromorphic solution of (\ref{eqKSODE}).
Notice that if a meromorphic solution $u$ has a pole at $a$, then $u(z+z_n-a)$ is also a meromorphic solution with a pole at $a$ because the ODE (\ref{eqKSODE}) is {\it autonomous}. Then a simple demonstration \cite{ConteNgODE3}, which essentially assumes the number of Laurent series to be finite
(which is our case), proves that any meromorphic solution must be periodic and hence elliptic or degenerate elliptic.

\begin{remark}
Using this method, one can prove that for a large class of $n$-th order autonomous algebraic ODEs, their meromorphic solutions must be elliptic or degenerate elliptic (see \cite{NW2018}). 
The method fails if there is a positive Fuchs index and hence one cannot conclude the finiteness of the number of Laurent series. 
In such cases, it is possible to have meromorphic solutions which are neither elliptic nor degenerate elliptic (see Lemma 4.5 of \cite{ConteNgWu2015}).  
\end{remark}

The remaining work is now to obtain all these solutions in closed form.

  \subsection{Hermite decomposition}
\label{sectionHermite_decomposition}

The classical decomposition of a rational function as a finite sum of poles
has been extended by Hermite to
rational functions of one exponential,
nondegenerate elliptic functions
and even to elliptic functions of the second and third kinds \cite{Hermite1873CoursX}.
It fits our goal (to find closed forms) for two reasons:
it is a finite expression and therefore closed form,
it makes use of the singularity structure established in Step 1.
Let us recall these results of Hermite,
a nice account of which can be found in a review by Paul Appell
\cite{Appell-Elements-simples}.
\medskip

Given a function $u(x)$ which is elliptic or degenerate elliptic,
its partial fraction decomposition 
(Ref.~\cite{Hermite-sum-zeta} in the elliptic case,
 Ref.~\cite[pages 2, 321, 352, 365]{Hermite1873CoursX} in the trigonometric or rational case)
is the sum of two parts:
the principle part 
(sum of the principal parts at the poles)
and the regular part (an entire function).
If one requires both parts to be elliptic or degenerate elliptic,
then the entire function can only be
a constant (elliptic case),
the sum of a polynomial of $e^{k x}$ and a polynomial of $e^{-k x}$ (rational of $e^{k x}$ case),
a polynomial of $x$ (rational case),
and the decomposition is unique,
\begin{eqnarray}
&&{\hskip -17.0 truemm}
\hbox{elliptic}:\ 
u=\left(\sum_{j=1}^N \sum_{q=0}^{n_j} c_{jq} \zeta^{(q)}(x-x_j)\right)+(\hbox{constant}),\
        \sum_{j=1}^N c_{j0}=0,\ 
				2 \le N,\
\label{eqDecomposition-elliptic}
\\ && {\hskip -17.0 truemm}
\hbox{rat}(e^{k x}),\! k\not\!=\!0:\!
u\!=\!\left(\!\sum_{j=1}^N\! \sum_{q=0}^{n_j} c_{jq} 
 \left(\!\frac{k}{2}\coth\frac{k}{2}(x-x_j)\!\right)^{\!\!(q)}\!\right)
\!\!+\!\! \left(\!\sum_{m=M_1}^{M_2}\! d_m (e^{k x})^m\!\right)\!\!,\!
0\! \le\! N, 
\label{eqDecomposition-trigo} 
\\ & &{\hskip -17.0 truemm}
\hbox{rat}(x):\
u=\left(\sum_{j=1}^N \sum_{q=0}^{n_j} c_{jq} (x-x_j)^{-q}\right)
+ \left(\sum_{m=0}^M d_m x^m \right),\
0 \le N,
\label{eqDecomposition-rational}
\end{eqnarray}
in which 
$\zeta$ is the function introduced by Weierstrass($\zeta'=-\wp$),
$N$ is the number of poles in a period parallelogram, 
$x_j$ the affixes of the poles,
$n_j$ the order of the pole $x_j$,
$M$ a positive integer,
$M_1$ and $M_2$ are integers,
$c_{jq}$, $d_m$, $k$ complex constants.

Note that each elementary unit (\textit{\'el\'ement simple})
has by convention one simple pole of residue unity at the origin.

The $c_{jq}$ coefficients are in one-to-one correspondence with
the coefficients of the principal part of the finitely many
Laurent series established in Step 1.

\subsection{Step 3. Nondegenerate elliptic solutions}

In our example (\ref{eqKSODE}),
since there is only one pole ($N=1$),
the constraint that the sum of residues of the Laurent series should vanish
imposes $b^2=16 \mu\nu$.
The closed form of Hermite is then
\begin{eqnarray}
& &
u=60 \nu \zeta'' + 15 b \zeta' + 0 \zeta + \hbox{constant},\
b^2=16 \mu \nu,\
\label{eqKSHermite-Elliptic}
\end{eqnarray}
and a straightforward computation yields (\ref{eqKSElliptic}).

\subsection{Step 4. Solutions rational in one exponential}

In the closed form (\ref{eqDecomposition-trigo}),
a simple computation
(we will come back to it later) first shows that the entire part
reduces to a constant,
yielding the closed form,
\begin{eqnarray}
u\!=\!60 \nu \tau''\! +\! 15 b \tau' 
 -\!\left(\!\frac{15 (b^2-16 \mu\nu)}{4 \times 19 \nu}\!+\!30 \nu k^2\!\right)\!
 \tau\! +\! \hbox{constant},\!
\tau\!=\!  \frac{k}{2} \coth \frac{k}{2}\, (\xi-\xi_0),
\label{eqKSHermite-Trigo}
\end{eqnarray}
and there only remains to compute $k^2$, the constant
and the possible constraints on the fixed parameters $\nu,b,\mu,A$.
The output is the six codimension-two solutions
(\ref{eqKSTrigo}) and Table \ref{TableKS}.

\subsection{Step 5. Rational solutions}

Finally,
the decomposition (\ref{eqDecomposition-rational})
of rational solutions reduces to
\begin{eqnarray}
& &
u = 120 \nu \chi^{-3} - 15b \chi^{-2}
        -\frac{15 (b^2-16 \mu \nu)}{4 \times 19 \nu} \chi^{-1}
        + \hbox{constant},\
\chi=x-x_0,
\end{eqnarray}
which yields the unique codimension-three solution (\ref{eqKSRational}).

\section{Lessons from this pedagogical example}

\subsection{On truncations}

\begin{enumerate}

\item
This example (\ref{eqKSODE}) is too simple 
in the sense that
truncation methods also succeed to achieve the full results.
Indeed, if one assumes the truncation
\cite{FournierSpiegelThual,KudryashovElliptic}
\begin{eqnarray}
& &
u= c_0 \wp' + c_1 \wp +c_2,\ c_0 \not=0,
\label{eqKSEllipticAssumption}
\end{eqnarray}
and eliminates the derivatives of $\wp$ with
\begin{eqnarray}
& &
\wp'' = 6 \wp^2 - \frac{g_2}{2},\
{\wp'}^2= 4 \wp^3 - g_2 \wp - g_3,
\end{eqnarray}
the left-hand side of (\ref{eqKSODE})
becomes an expression similar to (\ref{eqKSEllipticAssumption}),
i.e.~a polynomial in $\wp,\wp'$ of degree one in $\wp'$,
\begin{eqnarray}
& &
E(u)=E_{3,0} \wp^3 + E_{1,1} \wp \wp' + E_{2,0} \wp^2
   + E_{0,1} \wp'  + E_{1,0} \wp      + E_{0,0}=0,\
\end{eqnarray}
and the resolution of the six
\textit{determining equations} $E_{j,k}=0$
yields the unique solution (\ref{eqKSElliptic}).

Similarly, the trigonometric assumption
\begin{eqnarray}
& &
u=\sum_{j=0}^{-p} c_j \tau^{-j-p},\ c_0 \not=0,\ p=-3,\ q=-6,
\label{eqKSTruncation_tau}
\end{eqnarray}
if which $\tau$ is defined so as to have one simple pole of residue unity,
\begin{eqnarray}
& &
\tau' + \tau^2 + \frac{S}{2}=0,\
S=-\frac{k^2}{2}= \hbox{ constant} \in \mathbb{C},
\label{eqRiccatitau}
\end{eqnarray}
generates the seven determining equations
\begin{eqnarray}
& &
E(u) = \sum_{j=0}^{-q} E_j \tau^{-j-q}=0,\ \forall j:\ E_j=0,
\end{eqnarray}
whose solutions are the six trigonometric solutions 
(case $S\not=0$ and $\tau=\frac{k}{2} \coth \frac{k}{2}(x-x_0)$) 
and the rational solution (case $S=0$ and $\tau=1/(x-x_0)$).

\item
\textit{A contrario},
the example (\ref{eq_u_rational}), built to escape the truncation methods,
is easily integrated by the above procedure.
It admits two Laurent series with equal residues ($\chi=x-x_0$),
\begin{eqnarray}
& & {\hskip -16.0 truemm}
u = \frac{1}{6 \chi} +\varepsilon \frac{i \sqrt{2}}{24}
+\frac{35}{144} \chi
-\varepsilon \frac{3 i \sqrt{2}}{64} \chi^2+\dots,\
 \varepsilon=\pm 1,\
\ \hbox{Fuchs index } (-1),
\label{eqExample-rational-tanh}
\end{eqnarray}
it fits Clunie's lemma ($n=3, P=u, Q=-(1/36)({u'}^2 + 12 u^2 u'-(3/2) u'-(17/2) u^2+1/2$)),
and its two-pole Hermite decomposition with a constant entire part is,
\begin{eqnarray}
& & {\hskip -15.0 truemm}
u= \frac{1}{6} \left(\frac{k}{2} \coth \frac{k}{2} (x-x_1)
                    +\frac{k}{2} \coth \frac{k}{2} (x-x_2)\right)+c_0.
\label{eqExample-rational-tanh-Hermite}
\end{eqnarray}
One then identifies the Laurent expansion of (\ref{eqExample-rational-tanh-Hermite})
near $x=x_1$ (resp.~$x=x_2$)
to the Laurent series (\ref{eqExample-rational-tanh}) with $\varepsilon=1$
(resp.~$\varepsilon=-1$).
This yields $c_0=0, k^2=4,\ \tanh^2 ((x_1-x_2)/2)=-2$ or $-1/2$
(remember that $\tanh$ and $\coth$ only differ by a shift).

\end{enumerate}
\subsection{On positive integer Fuchs indices}

Let us explain on an elementary example why positive integer Fuchs indices 
harm the procedure.
Given the ODE with a meromorphic general solution \cite{CFP1993},
\begin{equation}
u'' + 3 u u' + u^3 = 0,\
 u=\frac{1}{x-a} + \frac{1}{x-b},\ a \hbox{ and } b \hbox{ arbitrary},
\label{eqTwopoles}
\end{equation}
the question is to integrate this ODE with the Hermite decomposition,
in the rational case to simplify.
Since it admits the two sets of Laurent series
($\chi$ denotes $x-x_0$, with $x_0$ movable),
\begin{eqnarray}
& & {\hskip -15.0 truemm}
\left\lbrace
\begin{array}{ll}
\displaystyle{
u = \chi^{-1} + \sum_{j=0}^{+\infty} (-1)^j a_1^{j+1} \chi^j,\ a_1 \hbox{ arbitrary},\
 \hbox{Fuchs indices } (-1,1),
}\\ \displaystyle{
u = 2 \chi^{-1},\  \hbox{Fuchs indices } (-2,-1),
}
\end{array}
\right.
\label{eqTwopoles-Laurent}
\end{eqnarray} 
the number of admissible rational Hermite decompositions is undetermined,
\begin{eqnarray}
& & {\hskip -15.0 truemm}
\left\lbrace
\begin{array}{ll}
\displaystyle{
u =\left(\sum_{j=1}^{N} \frac{1}{x-x_j}\right),\ 1 \le N,
}\\ \displaystyle{
u =\left(\sum_{j=1}^{N} \frac{1}{x-x_j}\right)+\frac{2}{x},\ 0 \le N,
}
\end{array}
\right.
\label{eqTwopoles-Hermite}
\end{eqnarray}
which is unpleasant, although one of these decompositions does succeed.
One way out is to look for a first integral of (\ref{eqTwopoles})
associated to the positive Fuchs index $1$.
Such a first integral $K$ does exist in the present case,
it is easily obtained with the assumption (\ref{eqsubeqODEOrderOnePP}) below with $m=2$, 
\begin{eqnarray}
& & {\hskip -15.0 truemm}
(u'+u^2)^2 + 4 K( 2 u' + u^2)=0,\ K= a_1^2,
\label{eqTwopolesFirst}
\end{eqnarray}
it admits finitely many (two) distinct Laurent series,
\begin{eqnarray}
& & {\hskip -15.0 truemm}
u = \chi^{-1} + \sum_{j=0}^{+\infty} (-1)^j (\pm \sqrt{K})^{j+1} \chi^j,\ 
 \hbox{Fuchs index } (-1),
\label{eqTwopolesFirst-Laurent}
\end{eqnarray}
and none of them admits the suppressed positive Fuchs index.
This defines two admissible Hermite decompositions,
\begin{eqnarray}
& & {\hskip -15.0 truemm}
u =\left(\sum_{j=1}^{N} \frac{1}{x-x_j}\right),\ N=1,2,
\label{eqTwopolesFirst-Hermite}
\end{eqnarray}
and one of them ($N=2$) succeeds to represent the general solution.

\begin{remark}
Once reduced to its first integral (\ref{eqTwopolesFirst}),
this example is similar to the previous one
(two Laurent series only).
\end{remark}


\section{Another characterization of elliptic solutions: the subequation method}

If one knows (or assumes) that the solutions
of the given $N$-th order ODE are elliptic,
the structure of its polar singularities allows one to characterize 
each such solution by a first order ODE,
which will still remain to be integrated.
This is the subequation method \cite{MC2003,CM2009},
based on two classical theorems of Briot and Bouquet, which we first recall.

\begin{definition}
Given an elliptic function,
its \textbf{elliptic order} is defined as
the number of poles in a period parallelogram,
counting multiplicity.
It is equal to the number of zeros.
\end{definition}

\begin{theorem} \cite[theorem XVII p.~277]{BriotBouquet}.
Given two elliptic functions $u,v$ with the same periods
of respective elliptic orders $m,n$,
they are linked by an algebraic equation
\begin{eqnarray}
& &
F(u,v) \equiv
 \sum_{k=0}^{m} \sum_{j=0}^{n} a_{j,k} u^j v^k=0,\ 
\label{eqTwoEllFunctions}
\end{eqnarray}
with $\Degree(F,u)=\Order(v)$, $\Degree(F,v)=\Order(u)$.
If in particular $v$ is the derivative of $u$,
the first order ODE obeyed by $u$ takes the precise form
\begin{eqnarray}
& &
F(u,u') \equiv
 \sum_{k=0}^{m} \sum_{j=0}^{2m-2k} a_{j,k} u^j {u'}^k=0,\ a_{0,m}\not=0.
\label{eqsubeqODEOrderOnePP}
\end{eqnarray}
\end{theorem}

\begin{theorem} (Briot and Bouquet \cite[vol II \S 139]{Valiron}).
If a first order $m$-th degree autonomous ODE has a singlevalued general solution,
\begin{itemize}
\item
it must have the form (\ref{eqsubeqODEOrderOnePP}),

\item
its general solution is either elliptic (two periods)
or rational in one exponential $e^{k x}$ (one period)
or rational in $x$ (no period)
(successive degeneracies $g_2^3-27 g_3^2=0$, then $g_2=0$ in
${\wp'}^2= 4 \wp^3 - g_2 \wp - g_3$).

\end{itemize}
\end{theorem}

\begin{remark}
Equation (\ref{eqsubeqODEOrderOnePP}) is invariant under an arbitrary
homographic transformation having constant coefficients,
this is another useful feature of elliptic equations.
\end{remark}

The algorithm of the subequation method is the following.

\textbf{Input}: an $N$-th order ($N\ge 2$) any degree autonomous algebraic
ODE 
\begin{eqnarray}
& &
E(u,u',\dots,u^{(N)})=0,\ '=\frac{\D}{\D x},
\label{eqODE}
\end{eqnarray}
admitting at least one Laurent series
\begin{eqnarray}
& &
u=\chi^p \sum_{j=0}^{+\infty} u_j \chi^j,\ \chi=x-x_0,\ -p \in \mathbb{N}.
\label{eqLaurent}
\end{eqnarray}

\textbf{Output}: all its elliptic or degenerate elliptic solutions
in closed form.

Successive steps \cite{MC2003,CMBook}:
\begin{enumerate}
\item
Enumerate the pole orders $m_i$ of all 
distinct Laurent series (excluding Taylor series).
This is Step 1 section \ref{sectionCounting}.
Deduce the list of elliptic orders $(m,n)$ of $(u,u')$,
with $m$ equal to all possible partial sums of the $m_i$'s.
For each element $(m,n)$ of the list, perform the next steps.

Example 1: the ODE (\ref{eqKSODE}) admits only one series $u$ with a triple pole, 
therefore $(m,n)$ can only be $(3,4)$.

Example 2: the ODE
\begin{eqnarray}
& & {\hskip -10.0 truemm}
E(u) \equiv \hbox{some differential consequence of }
a^2 {u'}^2 -(u^2 + b)^2 + c=0,\
\end{eqnarray}
admits at least two Laurent series ($m_i=1,n_i=2$, $i=1,2$),
\begin{eqnarray}
& &
u= \pm a \chi^{-1} + \dots,
\label{eqJacobiLaurent}
\end{eqnarray}
therefore the list of elliptic orders $(m,n)$ is $(1,2)$, $(2,4)$.

\item
Compute $J$ terms in the Laurent series,
with $J$ slightly greater than $(m+1)^2$.

\item
Define the first order $m$-th degree subequation $F(u,u')=0$ 
(it contains at most $(m+1)^2$ coefficients $a_{j,k}$),
\begin{eqnarray}
& &
F(u,u') \equiv
 \sum_{k=0}^{m} \sum_{j=0}^{2m-2k} a_{j,k} u^j {u'}^k=0,\ a_{0,m}\not=0.
\end{eqnarray}

\item
For each Laurent series (\ref{eqLaurent}) whose elliptic order $m_i$
contributes to the current sum $m$,
require the series to obey $F(u,u')=0$,
\begin{eqnarray}
& & {\hskip -10.0 truemm}
F \equiv \chi^{m(p-1)} \left(\sum_{j=0}^{\jmax} F_j \chi^j
 + {\mathcal O}(\chi^{\jmax+1})
\right),\
\forall j\ : \ F_j=0.
\label{eqLinearSystemFj}
\end{eqnarray}
and solve this \textbf{linear overdetermined} system for $a_{j,k}$.

\item
Integrate each resulting first order subequation $F(u,u')=0$.
This can be achieved either by the Hermite decomposition
(section \ref{sectionHermite_decomposition} above),
or section \ref{sectionRational-wp-wprime} below.

\end{enumerate}

\begin{remark}
The fourth step generates a \textit{linear},
infinitely overdetermined,
system of equations $F_j=0$ for the unknown finite set of coefficients $a_{j,k}$.
It is quite an easy task to solve such a system,
and this is the key advantage of the present algorithm.
\end{remark}


\subsection{Tutorial examples}

The travelling wave reduction of the Korteweg-de Vries (KdV) equation
\begin{eqnarray}
& &
u''' - \frac{6}{a} u u'=0,
\label{eqKdVreducelliptic}
\end{eqnarray}
admits an infinite number of Laurent series (notation $\chi=x-x_0$),
\begin{eqnarray}
& & {\hskip -10.0 truemm}
u=2 a \chi^{-2} + U_4 \chi^2 + U_6 \chi^4 + \frac{U_4^2}{6 a} \chi^6 + \dots,\
(U_4,U_6) \hbox{ arbitrary constants}.
\end{eqnarray}
In the infinite list $(m,n)=(2 k,3 k)$, $k \in \mathbb{N}$,
let us start with $k=1$ and define (step 3)
\begin{eqnarray}
& & {\hskip -10.0 truemm}
F \equiv
              {u'}^2
+ a_{0,1}     {u'}
+ a_{1,1} u   {u'}
+ a_{0,0}
+ a_{1,0} u
+ a_{2,0} u^2
+ a_{3,0} u^3, a_{0,2}=1.
\label{eqsubeqKdV}
\end{eqnarray}
This generates (step 4)
the linear overdetermined system (\ref{eqLinearSystemFj}),
\begin{eqnarray}
& &
\left\lbrace
\begin{array}{ll}
\displaystyle{
F_0 \equiv 16 a^2 a_{0,2} + 8 a^3 a_{3,0}=0,
}\\ \displaystyle{
F_1 \equiv -8 a^2 a_{1,1}=0,
}\\ \displaystyle{
F_2 \equiv 4 a^2 a_{2,0}=0,
}\\ \displaystyle{
F_3 \equiv -4 a a_{0,1}=0,
}\\ \displaystyle{
F_4 \equiv 2 a a_{1,0} -16 a a_{0,2} U_4 +12 a^2 a_{3,0} U_4=0,
}\\ \displaystyle{
F_5 \equiv 0,
}\\ \displaystyle{
F_6 \equiv a_{0,0} + 4 a a_{2,0} U_4 -32 a a_{0,2} U_6 +12 a^2 a_{3,0} U_6=0,
}\\ \displaystyle{
\dots}
\end{array}
\right.
\end{eqnarray}
whose unique solution is
\begin{eqnarray}
& &
{u'}^2 - (2/a) u^3 + 20 U_4 u + 56 a U_6=0.
\label{eqsubeqKdV0}
\end{eqnarray}
Therefore $U_4$ and $U_6$ are interpreted as the two
first integrals  of (\ref{eqKdVreducelliptic}),
and they are generated by the method.

Consider a second example which requires no computation,
\begin{eqnarray}
& &
E(u) \equiv 
a^2 {u'}^2 -(u^2 + b)^2 + c=0,
\end{eqnarray}
whose Laurent series are (\ref{eqJacobiLaurent}),
defining a list $(m,n)$ made of two elements, $(1,2)$ and $(2,4)$.
Step 4 generates the constraint $c=0$ for $(m,n)=(1,2)$ 
and no constraint for  $(m,n)=(2,4)$.
Step 5 yields the general solution in each case,
respectively 
a $\coth$ function $(c=0$)
and
a Jacobi elliptic function $(c\not=0$).

For the application to KS ODE (\ref{eqKSODE}),
see \cite{CM2009}.


\section{An alternative to the Hermite decomposition}
\label{sectionRational-wp-wprime}

This section only applies to first order autonomous equations 
which have the Painlev\'e property,
such as (\ref{eqsubeqODEOrderOnePP}),
and it represents its elliptic or degenerate elliptic solution
with a closed form different from that of Hermite.

A classical 19th century result due to Picard \cite[\S 33]{Valiron}
states that,
if an algebraic curve $F(u,v)=0$ can be parametrized by functions $u$ and $v$ meromorphic on $\mathbb{C}$,
then 
the genus can only be one or zero.

If the genus is one (nondegenerate elliptic case),
there exists a birational transformation between
$(u,v)$ and $(\wp,\wp')$,
\begin{equation}
   u=R_1 (\wp,\wp'),\
   v=R_2 (\wp,\wp'),\
\wp =R_3(u,v),\
\wp'=R_4(u,v),\
\label{eqbirawp}
\end{equation}
thus generating the representation (\ref{eqRationalwpwpprime})
of the elliptic function $u$.

If the genus is zero (degenerate elliptic),
the algebraic curve $F(u,v)=0$ admits a proper rational parametric representation,
\begin{equation}
u=R_1(t),\ v=R_2(t),\
\label{equnicursal}
\end{equation}
and the condition $\D u/ \D x=v$ yields for $t$ either
a rational function of one exponential $e^{k x}$
or a rational function of $x$.

Both cases are implemented 
in the package {\tt algcurves} written by Mark van Hoeij \cite{MapleAlgcurves} 
in the computer algebra language Maple.

The genus one case is processed as

{\tt with(algcurves);}     \hfill                     load the package 

{\tt genus} ((\ref{eqsubeqKdV0}),$u,u'$); \hfill     check that genus is one

{\tt Weierstrassform} ((\ref{eqsubeqKdV0}),$u,u',\wp,\wp'$); 
\hfill\break\noindent
the last command yielding the four formulae (\ref{eqbirawp}),
the first one being the desired answer.

In the genus zero case, the statement
{\tt parametrization} ((\ref{eqTwopolesFirst}),$u,u',t$); 
answers ($K=1$),
\begin{eqnarray}
& &
u=\frac{3 t^2+2 t-1}{2 t(t-1)},\
u'=\frac{\D u}{\D x}=-\frac{1+2 t^2+8 t^3+5 t^4}{4 t^2(1-t)^2}\cdot
\end{eqnarray}

\begin{remark}
The output (\ref{eqbirawp}) of {\tt Weierstrassform} is returned \textit{modulo}
the addition formula of $\wp$ and thus may not be simplified enough.
For instance,
{\tt Weierstrassform} ($u^4-u^3+u^2+u+4-{u'}^2,u,u',\wp,\wp'$,{\tt Weierstrass}) 
answers
\begin{eqnarray}
& & 
u=\frac{9 \wp-75-18 \wp'}{-143-6 \wp+9 \wp^2}\ccomma,\ g_2=-\frac{208}{3},\ g_3=-\frac{568}{27}\ccomma
\end{eqnarray}
while the simplest answer is $u$ independent of $\wp'$ and homographic in $\wp$.
\end{remark}


\section{The important case of amplitude equations}

Let us denote $\GLA(x,t)$ a complex amplitude                               
depending on the time $t$ and on one space variable $x$.
We consider two amplitude equations,
i.e.~partial differential equations obeyed by the complex amplitudes.
In the following,
$p$, $q$, $r$ denote complex constants,
and $\gamma$ a real constant. 
These equations are:
the one-dimensional cubic complex Gingburg-Landau equation (CGL3) 
\begin{eqnarray}
& &  {\hskip -18.0 truemm}
\hbox{(CGL3)}\
i \GLA_t +p \GLA_{xx} +q \mod{\GLA}^2 \GLA -i \gamma \GLA =0,\
p q \gamma \Im(q/p)\not=0,\
\label{eqCGL3}
\end{eqnarray}
the one-dimensional cubic-quintic complex Gingburg-Landau equation (CGL5) 
\begin{eqnarray}
& &  {\hskip -18.0 truemm}
\hbox{(CGL5)}\
i \GLA_t +p \GLA_{xx} +q \mod{\GLA}^2 \GLA +r \mod{\GLA}^4 \GLA -i \gamma \GLA =0,\
p r \gamma \Im(r/p)\not=0.
\label{eqCGL5}
\end{eqnarray}
We will also use the real notation
\begin{eqnarray}
& &  
\frac{q}{p}=d_r + i d_i,\ 
\frac{r}{p}=e_r + i e_i,\ 
\frac{1}{p}=s_r - i s_i.
\end{eqnarray}

These equations are generic equations for slowly varying amplitudes,
with many physical applications
(pattern formation, superconductivity, nonlinear optics, \dots),
see the reviews \cite{AK2002,vS2003}. 
In most physical applications,
the quintic term $r \mod{\GLA}^4 \GLA$ is zero.
Only when the cubic term fails to describe the required features 
(bifurcation, stability, etc) is the quintic term necessary.
We will only consider the most interesting (and most challenging)
physical situations,
i.e.~the so-called ``complex case'' in which 
$q/p$ (CGL3) or $r/p$ (CGL5) is not real.
Moreover, we will discard the plane wave solutions
\begin{eqnarray}
\GLA=\hbox{constant } e^{-i\omega t + i K x},
\label{eqCGL35Planewave}
\end{eqnarray}
because they don't capture the nonlinearity.

Since these two PDEs are autonomous, they admit a traveling wave reduction,
i.e.~a reduction to an ordinary differential equation (ODE)
in the independent variable $\xi=x-c t$, with $c$ an arbitrary real constant.

Physicists prefer to represent the complex amplitudes by their modulus and phase,
and therefore to define the traveling wave reduction of (\ref{eqCGL5})
by two real variables $M(\xi)$, $\varphi(\xi)$,
\begin{eqnarray}
& & {\hskip -15.0 truemm}
\left\lbrace
\begin{array}{ll}
\displaystyle{
\GLA(x,t) =\sqrt{M(\xi)} e^{ i(\displaystyle{-\omega t + \varphi(\xi)})},\
\xi=x-ct,\
c \hbox{ and }\omega \in \mathbb{R},\
}\\ \displaystyle{
 \frac{M''}{2 M} -\frac{{M'}^2}{4 M^2} + i \varphi''- {\varphi'}^2
        + i \varphi' \frac{M'}{M}
- i \frac{c}{2p} \frac{M'}{M}
+ \frac{c}{p} \varphi' 
+ \frac{q}{p} M
+ \frac{r}{p} M^2
+ \frac{\omega - i \gamma}{p}
=0.
}
\end{array}
\right.
\label{eqCGL35ReducMphi}
\label{eqCGL35SystemMphi}
\end{eqnarray}

{}From a mathematical point of view,
the search for solutions becomes much simpler
if one chooses a different representation for the complex amplitudes.
One first represents the traveling wave reduction
by one complex function $a$,
\begin{eqnarray}
& & {\hskip -15.0 truemm}
\left\lbrace
\begin{array}{ll}
\displaystyle{
          \GLA(x,t)=     \GLa (\xi) e^{-i \omega t},\
\overline{\GLA(x,t)}=\bar\GLa (\xi) e^{ i \omega t},\
c \hbox{ and }\omega \in \mathbb{R},\
}\\ \displaystyle{
p \GLa'' + q \GLa^2 \GLac + r \GLa^3 \GLac^2-i c \GLa' +(\omega - i \gamma) \GLa =0
\hbox{ and complex conjugate (c.c.)}
}
\end{array}
\right.
\label{eqCGL35Reduca}
\label{eqCGL35Systema}
\end{eqnarray}

Then, after introducing the logarithmic derivative,
\begin{eqnarray}
& &
(\log \GLa)'=\dloga,\
(\log \bar\GLa)'=-\dloga + (\log M)',\
\label{eqdlogadlogb}
\end{eqnarray}
the optimal system to be considered is
the closed system made of,
\begin{eqnarray}
& & {\hskip -15.0 truemm}
p (\dloga' + \dloga^2) + q M + r M^2 - i c \dloga +\omega- i \gamma =0,
\label{eqCGL35-mixture-Mod-dlog}
\end{eqnarray}
and its complex conjugate,
\begin{eqnarray}
& & {\hskip -15.0 truemm}
\left[
\barp \left(\frac{\D^2}{\D \xi^2} - 2 \dloga \frac{\D}{\D \xi}-\dloga'+\dloga^2 \right)
+ \barq M + \barr M^2 +i c \left( \frac{\D}{\D \xi} - \dloga\right) 
 + \omega+i \gamma
\right] M=0.
\label{eqCGL35-mixture-Mod-dlog-cc}
\label{eqCGL35-ODE-GLdla}
\end{eqnarray}

The further elimination of the modulus $M$ between
    (\ref{eqCGL35-mixture-Mod-dlog}) 
and (\ref{eqCGL35-mixture-Mod-dlog-cc})
defines
a single third order ODE for the logarithmic derivative $\dloga$.
We will not write explicitly this equation,
simply referring to it as (\ref{eqCGL35-ODE-GLdla}$\dloga$).

\begin{remark}
This single ODE (\ref{eqCGL35-ODE-GLdla}$\dloga$)
is a differential consequence of the Riccati ODE defined by setting $M=0$
in (\ref{eqCGL35-mixture-Mod-dlog}).
\end{remark}

\begin{remark}
For CGL3 and for the pure quintic case $q=0$ of CGL5, the degree of this third order ODE is one.
For the cubic-quintic case of CGL5 ($q\not=0$), 
because $M+q/(2 r)$ as defined by (\ref{eqCGL35-mixture-Mod-dlog})
is the square root of a differential polynomial of $\dloga$,
this degree is two,
therefore (\ref{eqCGL35-ODE-GLdla}$\dloga$)
may admit a singular solution 
(defined by canceling an odd-multiplicity factor of the discriminant);
this is indeed the case,
but this singular solution, defined by $M=0$, 
represents the plane wave solution (\ref{eqCGL35Planewave}),
which we have discarded.
\end{remark}

\begin{definition}
We call ``meromorphic traveling wave'' of CGL3/CGL5 
any solution in which either $M$ or $\varphi'$ or $\dloga$ 
is a function of $\xi$ meromorphic on $\mathbb{C}$.
\end{definition}

Step 2 (section \ref{sectionClunie}) of the method 
succeeds to prove that,
for both CGL3 and CGL5 
and for all values of the fixed parameters $p,q,r,\gamma,c,\omega$ of the reduced ODE,
all meromorphic traveling waves $M$ 
are elliptic or degenerate elliptic.
However, the proof 
(see Refs.~\cite{MC2003,Hone2005} for CGL3 
and Refs.~\cite{ConteNgCGL5_ACAP} for CGL5) 
needs a lot of details 
and for this reason will not be reproduced in this short chapter.
In particular,
the three functions $M$, $\varphi'$, $\dloga$ of $\xi$
are birationally equivalent \cite{ConteNgCGL5_ACAP},
therefore the meromorphy of anyone
implies the meromorphy of the two others.
As a consequence,
all meromorphic traveling waves $M$ of both CGL3 and CGL5 have been determined.
Several results had been previously obtained \cite{MC2003,Hone2005,VernovCGL5}
for finding such solutions, but they were incomplete.

Let us rather focus here on the following topics:
(i) the advantages of the logarithmic derivative $\dloga$ 
over the physical variables square modulus $M$ and phase $\varphi'$;
(ii) the singularity structure (Step 1);
(iii) the integration by the Hermite decomposition.

\subsection{Advantages of the logarithmic derivative variable}
\label{sectionTW-reduction-Logderiv}

There are two advantages.
The first one is that all the poles of the logarithmic derivative variable
are simple.

The second one is the following.
Suppose one has found an admissible Hermite decomposition for $\dloga$.
Then, by construction,
the corresponding complex amplitude $\GLA$ is recovered 
simply by one quadrature, see (\ref{eqdlogadlogb}),
\begin{eqnarray}
& & {\hskip -15.0 truemm}
\GLA=\GLA_0 e^{-i \omega t} e^{\int \hbox{(entire function part) }\D \xi}
     \prod_{j \in J} (\sigma(\xi-\xi_j))^{c_{j0}},
\label{eqAmplitude-Quadrature-A}
\end{eqnarray}
in which $\sigma(z)$ is the Weierstrass entire function or one of its degeneracies
$\sinh(k z)/k$ or $z$.

As opposed to previous work using the physical variables $M$ and $\varphi'$,
we therefore choose here to consider
the logarithmic derivative $\dloga$.


\subsection{Laurent series of $\dloga$ of CGL3 and CGL5}
\label{section-Laurent-CGL35}
\label{section-Laurent-CGL5}
\label{section-Laurent-CGL3}

As is well known, an elimination may create extraneous solutions,
therefore Laurent series computed 
from the single equation (\ref{eqCGL35-ODE-GLdla}$\dloga$)
may not all be admissible for
the system
(\ref{eqCGL35-mixture-Mod-dlog})--(\ref{eqCGL35-mixture-Mod-dlog-cc}).
To be safe, one should proceed as follows.

Using the single equation (\ref{eqCGL35-ODE-GLdla}$\dloga$),
one first computes all the polar behaviours of $\dloga$.
These poles are all simple,
the residues $\dloga_0$ being the three roots of
\begin{eqnarray}
& & {\hskip -10.0 truemm}
\hbox{(CGL3)}:\
(\dloga_0 - 1) \left\lbrack(\dloga_0+2)(\dloga_0 +3)\barp q - \dloga_0(\dloga_0-1)p \barq\right\rbrack=0,
\label{eqCGL3residues-logderiv}
\\ & & {\hskip -10.0 truemm}
\hbox{(CGL5)}:\ (\dloga_0-1)\left\lbrack(\dloga_0+1)(\dloga_0 +2)\barp r - \dloga_0(\dloga_0-1)p \barr\right\rbrack=0,
\label{eqCGL5residues-logderiv}
\end{eqnarray}
i.e.~the three values
\begin{eqnarray}
& & {\hskip -10.0 truemm}
\hbox{(CGL3)}:\ \dloga_0=1, -1 + i \alpha_1,\ -1 + i \alpha_2,\
\label{eqCGL3residues-logderiv-alpha}
\\ & & {\hskip -10.0 truemm}
\hbox{(CGL5)}:\ \dloga_0=1, -\frac{1}{2} + i \alpha_1,\ -\frac{1}{2} + i \alpha_2,
\label{eqCGL5residues-logderiv-alpha}
\end{eqnarray}
where $\alpha_1$, $\alpha_2$ are the two real roots defined by \cite{CT1989} \cite{MCC1994},
\begin{eqnarray}
& & {\hskip -12.0truemm}
\hbox{(CGL3)}:\ \alpha_k^2-3 \frac{d_r}{d_i} \alpha_k -2=0,\ d_i \not=0,
\label{eqAlphaBeta}
\\ & & {\hskip -12.0truemm}
\hbox{(CGL5)}:\ \alpha_k^2 -2 \frac{e_r}{e_i} \alpha_k -\frac{3}{4}=0,\  e_i \not=0.
\end{eqnarray}

Next, one computes the behaviour of $M$ when $\dloga \sim \dloga_0/(\xi-\xi_0)$,
by solving the algebraic first degree (CGL3 case) or second degree (CGL5 case) equation
(\ref{eqCGL35-mixture-Mod-dlog}) for $M$,
then by requiring $M$ to also obey (\ref{eqCGL35-mixture-Mod-dlog-cc}).
The behaviour of $(\log \bar\GLa)'$ follows by (\ref{eqdlogadlogb}),
and the result is summarized in Table \ref{TableLaurentCGL}.

\tabcolsep=0.5truemm

\begin{table}[h] 
\caption[Laurent series of $\dloga$.]{
         Laurent series of $\dloga$.
For each polar behaviour of $\dloga$,
this table displays:
the leading behaviours of $M$ and $(\log \bar\GLa)'$,
the Fuchs indices of the Laurent series of $\dloga$,
the number of distinct Laurent series of $\dloga$ of the considered line.
The (unphysical) solution $M=0$ of  (\ref{eqCGL35-mixture-Mod-dlog-cc})
is recovered by setting both arbitrary constants
$m_1$, $m_2$ to zero.
}
\vspace{0.2truecm}
\begin{center}
\begin{tabular}{| c | c | c | c | c | c | c |}
\hline 
     & Poles of $\dloga$ & $M$ & $(\log \bar\GLa)'$ & Indices & Nb($\dloga$) & Detail
\\ \hline \hline 
CGL3&$(-1+i\alpha_k)\xi^{-1}$&$\frac{3\alpha_k}{d_i}\xi^{-2}$&$(-1-i\alpha_k)\xi^{-1}$&$-1,\frac{7\pm\sqrt{1-24\alpha_k^2}}{2}$ & $2$ &
\\ \hline 
&$\xi^{-1}+\frac{i c}{2 p}$ & $m_1\xi+m_2\xi^2$ & $\frac{m_2}{m_1}-\frac{i c}{2 p}$ &$-1,3,4$& und.& 
\\ \hline \hline 
${\hbox{CGL5} \atop q\not=0}$& $\left(-\frac{1}{2}+i\alpha_k\right)\xi^{-1}$ & $\pm\sqrt{\frac{2\alpha_k}{e_i}}\xi^{-1}$ & $\left(-\frac{1}{2}-i\alpha_k\right)\xi^{-1}$ & $-1,\frac{5\pm\sqrt{1-32\alpha_k^2}}{2}$ & $4$ & (\ref{eqLaurent-CGL5-irrational})
\\ \hline 
&$\xi^{-1}+\frac{i c}{2 p}$ & $m_1\xi+m_2\xi^2$ & $\frac{m_2}{m_1}-\frac{i c}{2 p}$ &$-1,3,4$& und.& (\ref{eqLaurent-CGL5-residue-one})
\\ \hline \hline 
${\hbox{CGL5} \atop q=0}$& $\left(-\frac{1}{2}+i\alpha_k\right)\xi^{-1}$ & $\pm\sqrt{\frac{2\alpha_k}{e_i}}\xi^{-1}$ & $\left(-\frac{1}{2}-i\alpha_k\right)\xi^{-1}$ & $-1,\frac{5\pm\sqrt{1-32\alpha_k^2}}{2}$ & $2$ & (\ref{eqLaurent-CGL5-irrational})
\\ \hline 
		& $\xi^{-1}+\frac{i c}{2 p}$ & $m_1\xi+m_2\xi^2$ & $\frac{m_2}{m_1}-\frac{i c}{2 p}$ & $-1,4,5$ & und. & (\ref{eqLaurent-CGL5-residue-one})
\\ \hline \hline 
\end{tabular}
\end{center}
\label{TableLaurentCGL}
\end{table}

\begin{remark}
The table enumerating all the Laurent series of $M$ is quite different
from Table \ref{TableLaurentCGL}.
For instance, whatever be $q$ in CGL5, the poles of $M$ with principal parts 
$\pm\sqrt{\frac{2\alpha_k}{e_i}}\xi^{-1}$
define four distinct Laurent series of $M$,
and these are the only ones.
\end{remark}

For each of the two families with irrational Fuchs indices,
the number of Laurent series $\dloga$
is equal to the degree (one or two) of the third order algebraic ODE for $\dloga$
in its highest derivative $\dloga'''$.
For CGL3 and the case $q=0$ of CGL5,
this degree is one.
For the case $q\not=0$ of CGL5, this degree is two and
the two Laurent series only differ by signs,
\begin{eqnarray}
& & {\hskip -17.0 truemm}
\hbox{(CGL5)}\ \forall q:\ 
\dloga_0=-\frac{1}{2} + i \alpha_k,\ k=1,2,\
\nonumber\\ & & {\hskip -17.0 truemm} \phantom{123456789012}
\dloga= \dloga_0 \chi^{-1} 
+ i c \frac{(\dloga_0+1)(\dloga_0+3)}{8 p}
- i c \frac{\dloga_0^2-1}{8 \barp}
\nonumber\\ & & {\hskip -17.0 truemm} \phantom{12345678901234}
\pm \sqrt{\dloga_0(1-\dloga_0)\frac{p}{r}}
\left\lbrack 
-\frac{(\dloga_0+1)(\dloga_0+3)q}{8 p}+\frac{(\dloga_0-1) \barq}{8 \barp}
\right\rbrack + O(\chi).
\label{eqLaurent-CGL5-irrational}
\end{eqnarray}

For the residue unity of $\dloga$ in the CGL5 case,
the dependence on $m_1$, $m_2$ (the two arbitrary coefficients)
and $q$, $\barq$ of the couple $(\dloga,M)$ is,
\begin{eqnarray}
& & {\hskip -15.0 truemm}
\left\lbrace
\begin{array}{ll}
\displaystyle{
\dloga=\frac{1}{\xi}+\frac{i c}{2 p}+ . \xi -\frac{m_1 q}{4 p}\xi^2
+(.m_1^2+.m_2 q+.) \xi^3
+(.m_1 m_2 + . m_1 q + .m_2 q)\xi^4+O(\xi^5),
}\\ \displaystyle{
}\\ \displaystyle{
M=m_1 \xi + m_2 \xi^2 +(.m_1+.m_2) \xi^3
+(.m_1+.m_2+.m_1^2 q + .m_1^2 \barq) \xi^4 +O(\xi^5).
}
\end{array}
\right.
\label{eqLaurent-CGL5-residue-one}
\end{eqnarray}
in which the dots (.) represent constants independent of $m_1$, $m_2$, $q$, $\barq$.
This allows one to compute the Fuchs indices of $\dloga$
in both subcases $q=\barq=0$ and $q \barq\not=0$
and to check that the expansion is free from movable logarithms.

To conclude, the set of poles 
in the Hermite decomposition of
any solution $\dloga$ which is elliptic or degenerate elliptic is made of:
an undetermined number of poles of residue unity,
two poles (CGL3, resp.~CGL5 $q=0$) 
of residues $-1+ i \alpha_k$, resp.~$-1/2+ i \alpha_k$,
or 
four poles (CGL5 $q\not=0$) of residues $-1/2 + i \alpha_k$.
		
\subsection{Entire part of a toy ODE}
     \label{Entire_part_of_a_toy_ODE}
		
In the method, 
one must also compute the regular parts of the partial fraction decompositions,
Eqs.~(\ref{eqDecomposition-trigo}) and
     (\ref{eqDecomposition-rational}).		
Let us first present this computation	on a toy ODE.	
		
Let us consider (then forget) the rational function of $e^{k x}$,
\begin{eqnarray}
   & u & =k \coth k (x-x_0)/2 + a e^{k (x-x_0)} + b e^{-2 k (x-x_0)},
\label{eqToy-u}
\end{eqnarray}
build the first order autonomous ODE by elimination of $x_0$,
\begin{eqnarray}
   & & 2 {u'}^4 + \dots - 4 k^3 u^5=0,
\label{eqODEexercise1ux}
\end{eqnarray}
then process it 
in order to compute the Laurent polynomial 
$\sum_{m=-2}^1 d_m e^{m k x}$
(last two terms of (\ref{eqToy-u})).

\begin{enumerate}

\item 
Assuming $u$ to be a rational function of $X=e^{q x}$,
one builds the ODE for for $U(X)=u(x)$,
\begin{eqnarray}
   & & {\hskip -18.0 truemm} 
	  u=U,\ X=e^{q x},\
    E(U',U,X,q,k)\equiv 
    2 q^4 X^4 \left(\frac{\D U}{\D X}\right)^4 + \dots - 4 k^3 U^5=0.
\label{eqODEexercise1UXq}
\end{eqnarray}

\item 
Look for an algebraic behaviour 
$U \sim c X^p$, as $X \to \infty$, with $c p \not=0$,
\begin{eqnarray}
& & {\hskip -15.0 truemm}
X \to \infty,\ 
U \sim c X^p,\
E(U',U,X,q,k) \sim X^{5 p} c^5 (p q-k)(p q+2 k)^2,
\label{eqDecomposition-trigo-exercise-I}
\end{eqnarray}
which yields the two solutions,
\begin{eqnarray}
& & {\hskip -15.0 truemm}
\left\lbrace
\begin{array}{ll}
\displaystyle{
pq =   k,\ \ \ \ \ u \sim c e^{k x},\ \ \ \ c \hbox{ undetermined},
}\\ \displaystyle{
pq =-2 k,\         u \sim c e^{-2 k x},\    c \hbox{ undetermined},
}
\end{array}
\right.
\end{eqnarray}
and the final assumption for the Hermite decomposition of (\ref{eqODEexercise1ux})
\begin{eqnarray}
& & {\hskip -15.0 truemm}
u=q \coth \frac{q}{2}x+ \left(\sum_{m=-2}^{1} d_m (e^{k x})^m \right),\
\label{eqDecomposition-trigo-exercise}
\end{eqnarray}
with $q$ and the $d_m$'s to be determined.

\end{enumerate}

\subsection{Entire part of CGL3 and CGL5}
     \label{Entire_part_of_CGL3_and_CGL5}

Let us show that it reduces to a constant for both CGL3 and CGL5.

Following previous section \ref{Entire_part_of_a_toy_ODE},
one performs in (\ref{eqCGL35-ODE-GLdla}$\dloga$)
the change of function
\begin{eqnarray}
   & & (\dloga,\xi) \to (F,X):\ \dloga=F,\ X=e^{k \xi},\
	E(F''',F'',F',F,X,k) =0,
\label{eqODE-ekxi}
\end{eqnarray}
and one looks for power law behaviours of $F(X)$ when $X\to \infty$.
The result
\begin{eqnarray}
& &
X \to \infty,\
F \sim a X^m,\ m \not=0,\
\nonumber \\ & &
\hbox{(CGL3) } E(F''',F'',F',F,X,k) \sim a^4 p (\barp q - \barq p) X^{4 m},
\nonumber \\ & &
\hbox{(CGL5) } E(F''',F'',F',F,X,k) \sim a^{12} 2^6 p^4 r^3 (\barp r - \barr p)^2 X^{12 m},
\end{eqnarray}
yields no solution for $m k$
(as opposed to the toy ODE, see  Eq.~(\ref{eqDecomposition-trigo-exercise-I})),
therefore the regular part in the partial fraction decomposition
of the solution $\dloga$ of (\ref{eqCGL35-ODE-GLdla}$\dloga$)
reduces to a constant.

\subsection{All admissible partial fraction decompositions}
     \label{Admissible_partial_fraction_decompositions}

The information already obtained is
(i) the list of distinct Laurent series,
recalled in the end of section \ref{section-Laurent-CGL35};
(ii) the entire part of each decomposition,
which is constant.

Since all the poles of $\dloga$ are simple
(advantage of the logarithmic derivative),
the list of admissible Hermite decompositions of $\dloga$
is identical to the list of residues, counting multiplicity.
Let us prove that this list is finite.

Indeed, as proven in \cite{ConteNgCGL5_ACAP},
the number of Laurent series of $M$ is finite (and equal to $2$ or $4$),
therefore the number of admissible Hermite decompositions of $M$ is finite,
and so is the number of solutions $M$  
in the elliptic or degenerate elliptic class.

Since $\dloga$ is a rational function of $M$, $M'$, $M''$, $M'''$
(see details for instance in \cite{ConteNgCGL5_TMP}),
\begin{eqnarray}
& &
\dloga=\frac{M'}{M}+\frac{c s_r}{2} + \frac{G'-2 c s_i G}{2 M^2(g_r - d_i M - e_i M^2)},\
\label{eqCGL35Order3}
\\ & &
G=\frac{1}{2} M M'' - \frac{1}{4} M'^2
  -\frac{c s_i}{2} M M'  + g_i M^2 + d_r M^3 + e_r M^4,
\end{eqnarray}
the number of solutions $\dloga$ in that class is also finite,
and so is the number of their Hermite decompositions.
In particular, one concludes that 
the number $N_1$ of poles of $\dloga$ with residue unity
in any Hermite decomposition of $\dloga$ is finite,
and we leave it to the reader to establish the upper bound of $N_1$.

For CGL3, restricting to $N_1 \le 1$, 
the list contains six elements, characterized by their residues, 
\begin{eqnarray}
& & {\hskip -10.0 truemm}
\left\lbrace
\begin{array}{ll}
\displaystyle{
N=1:\ \lbrace \hbox{residues}\rbrace=\lbrace   -1+i \alpha_k             \rbrace,
}\\ \displaystyle{
}\\ \displaystyle{
N=2:\ \lbrace \hbox{residues}\rbrace=\lbrace 1,-1+i \alpha_k             \rbrace,
}\\ \displaystyle{
N=2:\ \lbrace \hbox{residues}\rbrace=\lbrace   -1+i \alpha_1,-1+i\alpha_2\rbrace,
}\\ \displaystyle{
}\\ \displaystyle{
N=3:\ \lbrace \hbox{residues}\rbrace=\lbrace 1,-1+i \alpha_1,-1+i\alpha_2\rbrace.
}
\end{array}
\right.
\label{eqCGL3-all-PFD}
\end{eqnarray}

The list for the pure quintic case $q=0$ of CGL5 is 
the same, after replacing $-1+i \alpha_k$ by  $-1/2+i \alpha_k$.

For the cubic-quintic case of CGL5 ($q\not=0$),
the multiplicity of each residue $-1/2+i \alpha_k$ is two,
thus defining a list made of twenty-four elements
(restricting to $N_1 \le 2$), 
\begin{eqnarray}
& & {\hskip -10.0 truemm}
\left\lbrace
\begin{array}{ll}
\displaystyle{
N=1:\ \lbrace \hbox{residues}\rbrace=\lbrace   -1/2+i \alpha_k             \rbrace,  
}\\ \displaystyle{
}\\ \displaystyle{
N=2:\ \lbrace \hbox{residues}\rbrace=\lbrace 1,-1/2+i \alpha_k             \rbrace,   
}\\ \displaystyle{
N=2:\ \lbrace \hbox{residues}\rbrace=\lbrace   -1/2+i \alpha_1,-1/2+i\alpha_2\rbrace,   
}\\ \displaystyle{
N=2:\ \lbrace \hbox{residues}\rbrace=\lbrace   -1/2+i \alpha_k,-1/2+i\alpha_k\rbrace,   
}\\ \displaystyle{
}\\ \displaystyle{
N=3:\ \lbrace \hbox{residues}\rbrace=\lbrace 1,1,-1/2+i \alpha_k\rbrace,   
}\\ \displaystyle{
N=3:\ \lbrace \hbox{residues}\rbrace=\lbrace 1,-1/2+i \alpha_1,-1/2+i\alpha_2\rbrace,   
}\\ \displaystyle{
N=3:\ \lbrace \hbox{residues}\rbrace=\lbrace 1,-1/2+i \alpha_k,-1/2+i\alpha_k\rbrace,   
}\\ \displaystyle{
N=3:\ \lbrace \hbox{residues}\rbrace=\lbrace   -1/2+i \alpha_1,-1/2+i \alpha_2,-1/2+i\alpha_k\rbrace,   
}\\ \displaystyle{
}\\ \displaystyle{
N=4:\ \lbrace \hbox{residues}\rbrace=\lbrace 1,1,-1/2+i \alpha_1,-1/2+i \alpha_2\rbrace,  
}\\ \displaystyle{
N=4:\ \lbrace \hbox{residues}\rbrace=\lbrace 1,1,-1/2+i \alpha_k,-1/2+i \alpha_k\rbrace,  
}\\ \displaystyle{
N=4:\ \lbrace \hbox{residues}\rbrace=\lbrace 1,-1/2+i \alpha_1,-1/2+i \alpha_2,-1/2+i\alpha_k\rbrace,  
}\\ \displaystyle{
N=4:\ \lbrace \hbox{residues}\rbrace=\lbrace   -1/2+i \alpha_1,-1/2+i \alpha_2,-1/2+i\alpha_1,-1/2+i \alpha_2\rbrace,  
}\\ \displaystyle{
}\\ \displaystyle{
N=5:\ \lbrace \hbox{residues}\rbrace=\lbrace 1,1,-1/2+i \alpha_1,-1/2+i \alpha_2,-1/2+i\alpha_k\rbrace,  
}\\ \displaystyle{
N=5:\ \lbrace \hbox{residues}\rbrace=\lbrace 1,-1/2+i \alpha_1,-1/2+i \alpha_2,-1/2+i\alpha_1,-1/2+i \alpha_2\rbrace,  
}\\ \displaystyle{
}\\ \displaystyle{
N=6:\ \lbrace \hbox{residues}\rbrace=\lbrace 1,1,-1/2+i \alpha_1,-1/2+i \alpha_2,-1/2+i\alpha_1,-1/2+i \alpha_2\rbrace.  
}
\end{array}
\right.
\label{eqCGL5qnonzero-all-PFD}
\end{eqnarray}

\section{Nondegenerate elliptic solutions}
\label{sectionNondegenerate_elliptic}

\subsection{Nondegenerate elliptic solutions of CGL3}

Given the set of three residues defined in (\ref{eqCGL3residues-logderiv-alpha}),
with respective multiplicities (undetermined, zero or one, zero or one),
the only realization of the constraint $\sum_{j=1}^N c_{j0}=0$
in (\ref{eqDecomposition-elliptic})
is: two copies of the Laurent series with residue one,
plus one copy of each of the two others,
with the additional condition $\alpha_1=-\alpha_2=\sqrt{2}$ (i.e.~$d_r=0$).
However,
the zero sum condition further applied to various differential polynomials of $\dloga$
(which must also be elliptic)
generates additional necessary conditions
which prevent CGL3 to admit nondegenerate elliptic solutions.

This result was first established by Hone \cite{Hone2005} 
by considering the variable $\mod{\GLA}^2$,
but his argument is here simplified by the proof that each residue $-1+i \alpha_k$
is associated to one and only one Laurent series.

\subsection{Nondegenerate elliptic solutions of CGL5}

Similarly,
given the three residues (\ref{eqCGL5residues-logderiv-alpha}),
with respective multiplicities (unknown, zero or one or two, zero or one or two),
there exist two sets whose weighted sum is zero,
obtained by taking each residue once or twice
with the additional condition $\alpha_1=-\alpha_2=\sqrt{2}$ (i.e.~$e_r=0$),
\begin{eqnarray}
& &
{\hskip -8 truemm}
\hbox{first set, any }q:\
1,\
\frac{-1+i \sqrt{3}}{2},\
\frac{-1-i \sqrt{3}}{2},
\\  & & 
{\hskip -8 truemm}
\hbox{second set }
(q \not=0):\
1,\
1,\
\frac{-1+i \sqrt{3}}{2},\
\frac{-1+i \sqrt{3}}{2},\
\frac{-1-i \sqrt{3}}{2},\
\frac{-1-i \sqrt{3}}{2}\cdot
\end{eqnarray}

For the first set,
the zero sum condition further applied to various differential polynomials of $\dloga$
generates three more necessary conditions,
equivalent to the conditions \cite[Eq.~(25)]{ConteNgCGL5_TMP}
generated by the consideration of $M$ instead of $\dloga$.
The result is a unique elliptic solution,
\begin{eqnarray}
& & {\hskip -18.0 truemm}
\left\lbrace {\hskip -1.8 truemm}
\begin{array}{ll}
\displaystyle{
q=0,\ \Re\left(\frac{r}{p}\right)=0,\
\Im\left(\frac{\gamma + i \omega}{p}\right) = -\frac{1}{4} (c s_r)^2-\frac{3}{16} (c s_i)^2,\
}\\ \displaystyle{
\frac{\D}{\D \xi} \log\! \left(\!\GLA e^{\displaystyle{i \omega t\! -\! i \frac{c s_r}{2} \xi}}\!\right)
\!=\!
\frac{c s_i}{2}
\!+\!\sum\limits_{k=1}^3
  e^{k 2 i \pi/3}
	\left(\zeta(\xi\!-\!\xi_{j,k}^{\dloga},G_2,G_3) \!+\!\zeta(\xi_{j,k}^{\dloga},G_2,G_3)\right)\!,
%
}\\ 
\displaystyle{
M=\frac{3^{1/4}}{\sqrt{-e_i}} \sum_{k=1}^4
  e^{k 2 i \pi/4}\left(\zeta(\xi-\xi_{j,k}^{M},g_2,g_3) +\zeta(\xi_{j,k}^{M},g_2,g_3)\right).
}
\end{array}
\right.
\label{eqCGL5-Elliptic-sol-three-poles}
\end{eqnarray}
Detailed in \cite[Eq.~(61)]{ConteNgCGL5_ACAP},
this codimension-four solution is an extrapolation of a previous result \cite{VernovCGL5},
and the relation $M=\mod{\GLA}^2$ defines a Landen transformation
\cite[Appendix]{ConteNgCGL5_ACAP}
between $\zeta(*,g_2,g_3)$ and $\zeta(*,G_2,G_3)$.

\begin{remark}
Out of the three $\zeta$ functions in the expression 
(\ref{eqCGL5-Elliptic-sol-three-poles}) for $\dloga$,
two arise from the Laurent series (\ref{eqCGL5residues-logderiv-alpha})
with irrational Fuchs indices
and one from the Laurent with positive integer indices.
This is in contrast with the expression for $M$,
made of four Laurent series with irrational Fuchs indices
(see \cite[Eq.~(21)]{ConteNgCGL5_ACAP}),
without any contribution from some Laurent series with positive integer indices.
\end{remark}

As to the second set, it cannot define a nondegenerate elliptic solution
since, as proven in \cite{ConteNgCGL5_ACAP},
(\ref{eqCGL5-Elliptic-sol-three-poles}) is the unique such solution.

\section{Degenerate elliptic solutions}
\label{sectionDegenerate-elliptic-solutions}

\subsection{Method of resolution} 
     \label{Method_of_resolution} 

Processing the single third order ODE (\ref{eqCGL35-ODE-GLdla}$\dloga$)
might result in too large expressions,
therefore it is better to again take advantage of the logarithmic derivative,
as we now illustrate on the example of
trigonometric solutions of CGL3 made of one principal part.

Given the trigonometric one-pole assumption
\begin{eqnarray}
& &
\dloga=\CA+ \dloga_0 \frac{k}{2} \coth \frac{k}{2} \xi,\ k\not=0,\ \dloga_0 \not=1,\
\label{eqCGL3-dloga-One-pole}
\end{eqnarray}
one quadrature yields
\begin{eqnarray}
& &
\GLa=\GLA_0 e^{\CA \xi} \left(\sinh \frac{k}{2} \xi\right)^{\dloga_0},\
\GLA_0=\hbox{ arbitrary complex constant},\
\end{eqnarray}
and, by complex conjugation (using the property $\Re(\dloga_0)=-1$),
\begin{eqnarray}
& &
\bar \GLa=\bar\GLA_0 e^{\bar\CA \xi} \left(\sinh \frac{k}{2} \xi\right)^{-2-\dloga_0},\
M=m_0 \ e^{(\CA+\bar\CA)\xi} \ \frac{k^2}{4} \left(\coth^2 \frac{k}{2} \xi - 1\right),
\end{eqnarray}
therefore the constant $\CA$ must be purely imaginary.

The above values of $\dloga$ and $M$ inserted in 
(\ref{eqCGL35-mixture-Mod-dlog}) and
(\ref{eqCGL35-mixture-Mod-dlog-cc})
generate two polynomials, 
\begin{eqnarray}
& & {\hskip -10.0 truemm}
(\ref{eqCGL35-mixture-Mod-dlog})   =\sum_{j=0}^2  E_j z^j,\ 
(\ref{eqCGL35-mixture-Mod-dlog-cc})=\sum_{j=0}^4 E'_j z^j,\ z=\frac{k}{2}\coth \frac{k}{2} \xi.
\end{eqnarray}
Requiring that they identically vanish defines eight determining equations
in the unknowns ($\dloga_0$, $m_0$, $\CA$, $\omega$, $k^2$, $c$),
and the parameters ($p$, $\barp$, $q$, $\barq$, $\gamma$),
\begin{eqnarray}
& & {\hskip -10.0 truemm}
\left\lbrace
\begin{array}{ll}
\displaystyle{
E_2 \equiv q m_0 + p \dloga_0 (\dloga_0-1)=0,
}\\ \displaystyle{
E_1 \equiv \dloga_0 (2 p \CA - i c)=0,
}\\ \displaystyle{
E_0 \equiv -4 i c \CA -4 i \gamma + 4 \omega + 4 p \CA^2 + (p \dloga_0 - q m_0) k^2=0, 
}
\end{array}
\right.
\end{eqnarray}
and
\begin{eqnarray}
& & {\hskip -17.0 truemm}
\left\lbrace {\hskip -1.5 truemm}
\begin{array}{ll}
\displaystyle{
E'_4 \equiv (\barq m_0 + \barp (\dloga_0+2) (\dloga_0+3)) m_0=0,
}\\ \displaystyle{
E'_3 \equiv (\dloga_0 +2)(2 \barp \CA- i c) m_0=0,
}\\ \displaystyle{
E'_2 \equiv \left[-4 i c \CA+4 i\gamma+4\omega+4\barp\CA^2+(-\barp (\dloga_0+2)(\dloga_0+4)- 2 \barq m_0) k^2\right] m_0=0,
}\\ \displaystyle{
E'_1 \equiv k^2 E'_3=0,
}\\ \displaystyle{
E'_0 \equiv k^4 E'_4 - k^2 E'_2=0.
}
\end{array}
\right.
\end{eqnarray}

\medskip

Two kinds of resolutions can be performed.

The first one is to solve the three equations $E_j=0$ 
(which are linearly independent) on the field $\mathbb{C}$
as a \textit{linear} system of Cramer type (for instance in $m_0$, $\CA$, $\omega$),
then to require that $m_0$, $i \CA$, $\omega$, $k^2$ and $c$ be real.
These reality conditions ensure by construction
that the other five equations $E'_j=0$ are satisfied. 

The second one is to solve the eight equations $E_j=0$, $E'_j=0$
(which are no more independent) again on the field $\mathbb{C}$,
as an overdetermined \textit{linear} system for some set of variables,
be they unknowns or parameters (this makes no difference,
as argued in \cite[Appendix A]{CM1993}).
As usual, equations should be processed by decreasing values of 
the singularity degree $j$.
Such a set is, for instance,
$(m_0, \barq)$, then $(\CA,c)$, then $(\omega,\gamma)$.
After that,
there is no need to enforce any reality condition
since they are already taken into account.
\medskip

Since the residue $\dloga_0=1$ has been excluded in assumption (\ref{eqCGL3-dloga-One-pole}),
the solution is unique, 
\begin{eqnarray}
& & {\hskip -15.0 truemm}
\left\lbrace
\begin{array}{ll}
\displaystyle{
m_0=-\dloga_0(\dloga_0-1)\frac{p}{q},\
\barq=\frac{(\dloga_0+2)(\dloga_0+3)}{\dloga_0(\dloga_0-1)}\frac{\barp q}{p},\
\CA=i \frac{c}{2 p},\ 
c(p-\barp)=0,\
}\\ \displaystyle{
\omega =  -\frac{k^2}{8} (p \dloga_0^2+\barp(\dloga_0+2)^2)- \frac{c^2}{4 p},\
\gamma = -i\frac{k^2}{8} (p \dloga_0^2-\barp(\dloga_0+2)^2),\
}
\end{array}
\right.
\end{eqnarray}
or equivalently in the physical variables,
\begin{eqnarray}
& & {\hskip -15.0 truemm}
\left\lbrace
\begin{array}{ll}
\displaystyle{
\GLA= \GLA_0 e^{-i \omega t + i c \xi/(2 p)} \frac{k}{2} \left(\sinh \frac{k}{2} \xi\right)^{-1+i \alpha},\
M=\mod{\GLA_0}^2 \frac{k^2}{4} \left(\coth^2 \frac{k}{2} \xi - 1\right),\
}\\ \displaystyle{
\mod{\GLA_0}^2=\frac{3\alpha}{d_i},\
\frac{i \gamma-\omega}{p}=\left(\frac{c}{2 p}\right)^2 + \frac{k^2}{4}(1-i \alpha)^2,\
\Im(c/p)=0,\
}
\end{array}
\right.
\end{eqnarray}
which splits in two solutions.
The case $c=0$ is the stationary pulse or solitary wave \cite{PS1977},
\begin{eqnarray}
& & {\hskip -15.0 truemm}
\left\lbrace
\begin{array}{ll}
\displaystyle{
\GLA=\GLA_0 e^{-i\omega t}
\left(\cosh K x \right)^{-1+i \alpha},\
\frac{i \gamma- \omega}{p}=(1-i \alpha)^2 K^2,\
c=0,\
\lim_{x\to -\infty} \mod{\GLA}=\lim_{x\to +\infty} \mod{\GLA}=0,
}\\ \displaystyle{
c=0,\
\mod{\GLA_0}^2=\frac{3\alpha}{d_i},\
\frac{\gamma} {2 \alpha s_r +(\alpha^2-1)s_i}
=\frac{\omega}{2 \alpha s_i -(\alpha^2-1)s_r}
=-\frac{k^2}{4 (s_r^2+s_i^2)},
}
\end{array}
\right.
\label{eqCGL3Pulse}
\end{eqnarray}
and the case $\Im(1/p)=0$ is,
\begin{eqnarray}
& & {\hskip -15.0 truemm}
s_i=0,\
\mod{\GLA_0}^2=\frac{3\alpha}{d_i},\
k^2=-\frac{2 s_r \gamma}{\alpha},\
\omega=\frac{2 (1-\alpha^2)\gamma - \alpha s_r c^2}{4 \alpha}\ccomma\
c \hbox{ arbitrary}.
\end{eqnarray}

\subsection{Results} 
     \label{CGL35Results} 

   By lack of space, we postpone the exhaustive list of results to a forthcoming paper,
but, essentially, the only new recent result is the elliptic solution
of the purely quintic CGL5 \cite[Eq.~(61)]{ConteNgCGL5_ACAP},
recalled in Eq.~(\ref{eqCGL5-Elliptic-sol-three-poles}). 	

\section{Current challenges and open problems}

The two examples of this chapter are not independent,
and CGL3 admits a scaling limit \cite{PM1979}
under which the variable $u=\arg A$ obeys the KS PDE (\ref{eqKS}),
this is why in this section we concentrate on KS.

\begin{figure}[h]
\begin{center}
\epsfig{file=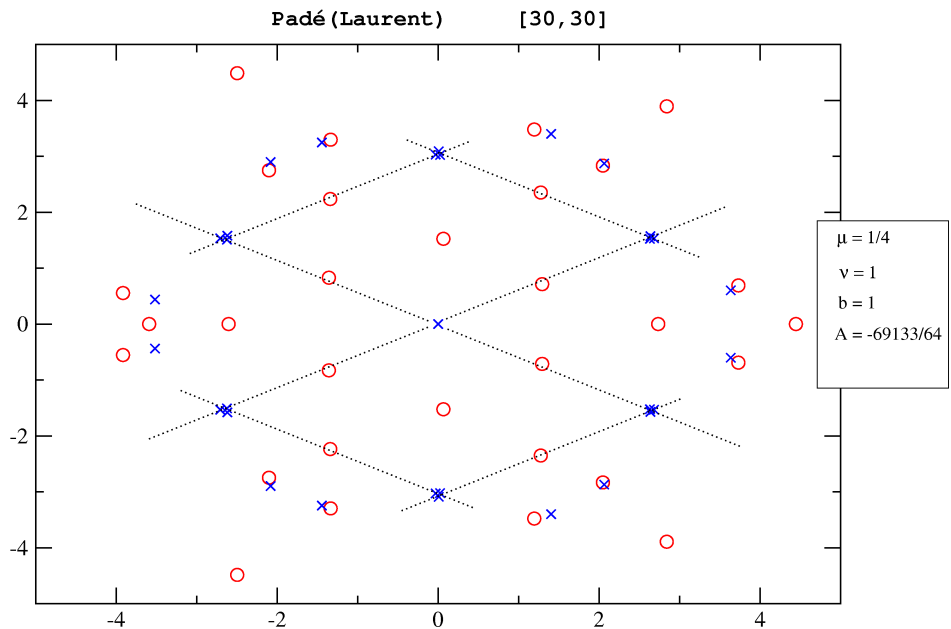,width=0.90\textwidth}
\end{center}
\caption[KS. Pattern of singularities of the unknowm general analytic solution
]
{Singularities of the Pad\'e approximant [L/M]
of the unknown solution of KS ODE \ref{eqKSODE}
(courtesy of Tony Yee Tat-leung).
Crosses are poles, circles zeroes.
The numerical values are
$L=30,M=30,\nu=1,b=1,\mu=1/4,A=-69133/64$.
}
\label{FigKSPade}
\end{figure}

\begin{enumerate}
\item
All meromorphic solutions of the third order ODE (\ref{eqKSODE})
have been found.

\item
Since they all have a nonzero codimension,
the problem remains open to find a closed form of the Laurent series
(\ref{eqKSODELaurent}) for generic values of $(\nu,b,\mu,A)$.
If it exists,  
the results of Eremenko prove that it is not meromorphic.

\item
A numerical investigation by Pad\'e approximants \cite{YCM2003} 
of the singularities of the
sum of the Laurent series (\ref{eqKSODELaurent})
for generic values of $(\nu,b,\mu,A)$
confirms (this is not a proof) the absence of any multivaluedness
and displays a nearly doubly periodic pattern for the singularities,
the unit cell being made of one triple pole and three simple zeroes
(Fig.~\ref{FigKSPade}).
Then singlevalued nonmeromorphic closed forms matching this description
could involve the expression of Picard \cite{Picard1893}
$\wp((\omega/(i \pi)) \log (x-c_1)+c_2,g_2,g_3)$ (with $2 \omega$ a period of $\wp$),
which admits the isolated movable noncritical essential singularities 
$x=c_1+2 n i \pi$, $n$ integer.

\end{enumerate}

\begin{acknowledgments}
This work was partially funded by the 
Hong Kong
RGC grant 17301115.
The first author is grateful to the Institute of mathematical research of HKU
for financial support. 
\end{acknowledgments}


\vfill\eject
\end{document}